\documentclass[preprint,showpacs,preprintnumbers,amsmath,amssymb,showkeys]{revtex4}
 
\usepackage{graphicx}% Include figure files
\usepackage{dcolumn}% Align table columns on decimal point
\usepackage{bm}% bold math

\begin{document}

\noindent

\preprint{}

\title{Quantum fluctuations from a local-causal information dynamics}% Force line breaks with \\
 
\author{Agung Budiyono}  
\email{agungbymlati@gmail.com} 

\affiliation{Jalan Emas 772 Growong Lor RT 04 RW 02 Juwana, Pati, 59185 Jawa Tengah, Indonesia}

\date{\today}% It is always \today, today,
             %  but any date may be explicitly specified
 
\begin{abstract} 
We shall show that the abstract and formal rules which govern the quantum kinematic and dynamics can be derived from a law of change of the information content or the degree of  uncertainty that the system has a certain configuration in a microscopic time scale, which is singled out uniquely, up to a free parameter, by imposing the condition of Macroscopic Classicality and the principle of Locality. Unlike standard quantum mechanics, however, the system always has a definite configuration all the time as in classical mechanics, following a continuous trajectory fluctuating randomly in time.    
\end{abstract}  

% PACS, the Physics and Astronomy
                             % Classification Scheme.
%Use showkeys class option if keyword
                              %dSsplay desired
\maketitle 

\section{Motivation} 

The violation of Bell inequality by quantum mechanics is widely believed to lead to a bizarre conclusion that quantum mechanics allows the statistical results of a pair of measurement events spacelike separated from each other to have a stronger correlation than that is allowed by any `local-causal' theory \cite{Bell paper,CHSH inequality,Bell book}. The `nonlocal correlation' has been claimed to be verified in numerous experiments \cite{Clauser experiment,Aspect experiment,Mandel experiment,Alley experiment,Tapster experiment,Ou experiment,Martienssen experiment,Kwiat experiment,Weihs experiment,Gisin experiment,Rowe experiment,Monroe experiment,Salart experiment}, in spite of the fact that no experiment hitherto conducted is free from loopholes \cite{Santos loophole,Brunner review}. Such a nonlocality prima facie contradicts the spirit of the  special theory of relativity which presumes a finite maximum velocity of interaction given by the velocity of light in vacuum. Further careful investigation however showed that the quantum mechanical nonlocal correlation can not be exploited by one party to influence the statistical results of measurement performed by the other distantly separated party, thus prohibits signaling, in accord with the assertion of the special theory of relativity \cite{Eberhard no-signaling,GRW no-signaling,Page no-signaling,Jarret no-signaling,Shimony 0}.  
    
The co-existence of nonlocal correlation and no-signaling in quantum mechanics has inspired some authors to ask if quantum mechanics can be derived from a certain balance between some kind of nonlocality and the principle of no-signaling \cite{Shimony 1,Shimony 2,Popescu no-signaling 1,Grunhaus no-signaling,Popescu no-signaling 2}. While it is shown that the constraints put by the nonlocal correlation and no-signaling are {\it not} sufficiently strong to single out quantum mechanics \cite{Popescu no-signaling 1,Grunhaus no-signaling,Popescu no-signaling 2}, it has renewed an interest in an approach to clarify the meaning of quantum mechanics by deriving its formal mathematical structures and numerous abstract postulates from a set of conceptually simple and physically transparent axioms. In such a program, one attempts to directly answer the most tantalizing foundational question: ``why the quantum?'' \cite{Wheeler}.  One of the advantages of the program to reconstruct quantum mechanics is that it might provide physical insights for possible natural extensions of quantum mechanics either by modifying the axioms or varying the free parameters that are left unfixed by the axioms. Extension of quantum mechanics is not only necessary to set up precision tests against quantum mechanics, but might turn out to be the necessary step to solve some of the foundational problems of the latter.   

A lot of works along this line has been reported recently by regarding `information' as the basic ingredient of Natural phenomena \cite{Rovelli,Zeilinger information quantization,Hardy construction,Simon,Clifton,van Dam,Brassard,Pawlowski,Goyal,Oppenheim,Barnum,Navascues,Brukner2,Masanes,Chiribella,Caticha,Torre,Fivel} : ``all things physical is information-theoretic in origin'' thus ``It from Bit'' \cite{Wheeler}. In those works, one searches for a set of basic features of {\it information processing} which can be promoted as axioms to reconstruct quantum mechanics. Such an approach is partly motivated by the advancement of quantum information science \cite{QI}: the fact that quantum mechanics allows information processing tasks that can not be performed at least as efficiently within classical mechanics suggests an intimate relationship between the the foundation of quantum mechanics and the basic features of information processing. This approach thus plays with information within operational-instrumentalist theoretical framework in which the notion of preparation and measurement play central role. Another different way to reconstruct quantum mechanics is to assume that quantum fluctuations is objectively real thus should be properly modeled by a stochastic processes. A lot of efforts have been made within this realist theoretical framework to derive the Schr\"odinger equation from a stochastic processes \cite{Fenyes,Weizel,Kershaw,Nelson,dela Pena 1,Davidson,dela Pena 2,Blanchard,Guerra,Garbaczewski,dela Pena 3,Markopoulou,Santos,dela Pena 4}. The greatest challenge of such an approach is how to explain the nonlocal correlation widely believed to be a feature of quantum mechanics.  

In the present paper, we shall follow the above second point of view. We shall first propose a statistical model of stochastic deviation from classical mechanics in microscopic regime based on a stochastic fluctuations of infinitesimal stationary action. We shall then show that the abstract and ``strange'' \cite{Giulini strange} rules of quantization of classical systems can be derived from a specific law of infinitesimal change of the `information content' or the `degree of uncertainty' that `the system {\it has} a certain configuration' along an infinitesimally short path, induced by the stochastic fluctuations of the infinitesimal stationary action. This law for the dynamics of information is shown to be singled out uniquely, up to a free parameter, by imposing the condition of Macroscopic Classicality and the principle of Local-Causality. Note that here, as will be detailed later, information is used to quantify an {\it actual} degree of uncertainty referring directly to an event regardless of measurement. It is then imperative to ask: how to explain the violation of Bell inequality in experiments? Putting the problem aside, we will show that the local-causal statistical model thus developed leads to the derivation the linear Schr\"odinger equation with Born's statistical interpretation of wave function and quantum mechanical uncertainty relation, two of the cornerstones of standard quantum mechanics.     We shall thus argue that quantization is physical and Planck constant acquires physical interpretation as a statistical average of the stochastic deviation from classical mechanics in a microscopic time scale. Moreover, unlike the standard canonical quantization, the system always has a definite configuration all the time as in classical mechanics, fluctuating randomly with time.   

\section{A statistical model of microscopic randomness and the dynamics of uncertainty obeying the principle of Locality} 

\subsection{A class of statistical models of microscopic stochastic deviation from classical mechanics}

There is a wealth of empirical evidences that phenomena in microscopic regime involve a universal stochastic element. Yet, unlike the Brownian motion, hitherto there is no consensus on the nature and origin of its randomness. Moreover, the prediction of quantum mechanics on the AB (Aharonov-Bohm) effect \cite{Aharonov-Bohm} and its experimental verification \cite{Peshkin} suggest that the randomness in microscopic regime can not be adequately described by introducing some kinds of conventional random forces as in Brownian motion. The force has to act at a distance. 

To discuss the universal randomness in microscopic regime, let us consider the following class of statistical models. Let $q$ denotes the configuration of the system and $t$ is time parameterizing the evolution of the system. Let us assume that the Lagrangian is parameterized by a random variable $\xi$ fluctuating in a microscopic time scale $dt$, whose origin is not our present concern: $L=L(q,\dot{q};\xi)$, where $\dot{q}\doteq dq/dt$. Let us then consider two infinitesimally close spacetime points $(q;t)$ and $(q+dq;t+dt)$ such that $\xi$ is constant. Let us assume that fixing $\xi$, the principle of stationary action is valid to select a path, denoted by $\mathcal{J}(\xi)$, that connects the two points. One must then solve a variational problem $\delta(L dt)=0$ with fixed end points. This leads to the existence of a function, the Hamilton's principal function denoted by $A(q;t,\xi)$, whose differential along the segment of trajectory is given by \cite{Rund book}, for a fixed $\xi$, 
\begin{equation}  
dA=Ldt=p\cdot dq-Hdt, 
\label{infinitesimal stationary action}
\end{equation} 
where $p(\dot{q})=\partial L/\partial{\dot{q}}$ is the classical momentum and $H(q,p)\doteq p\cdot\dot{q}(p)-L(q,\dot{q}(p))$ is the classical Hamiltonian. The above relation implies the following Hamilton-Jacobi equation:
\begin{eqnarray}
p=\partial_qA,\hspace{9mm}\nonumber\\
-H(q,p)=\partial_tA.  
\label{Hamilton-Jacobi condition}
\end{eqnarray}   

Varying $\xi$, the principle of stationary action will therefore pick up various different paths $\mathcal{J}(\xi)$, all connecting the same two infinitesimally close spacetime points, each might have different values of infinitesimal stationary action $dA(\xi)$. $dA(\xi)$ thus is randomly fluctuating due to the random fluctuations of $\xi$. The system starting with a configuration $q$ at time $t$ may therefore take various different paths randomly to end up with a configuration $q+dq$ at time $t+dt$. We have thus a stochastic processes driven by the random fluctuations of $\xi$ in a microscopic time scale. Hence a complete description of a single event is impossible. Instead, one has to rely on a statistical approach. 

One can see that the randomness enters into the dynamics in a microscopic time scale in a fundamentally different way from that of the Brownian motion. In the model, it is the infinitesimal stationary action that is randomly fluctuating in a microscopic time scale. By contrast, the randomness in the Brownian motion is induced by some random forces. We have thus assumed that the Lagrangian schema based on energies is more fundamental than the Newtonian schema based on forces. We expect that this will lead to a local-causal explanation of the AB effect. To see another implication of such a difference, let us consider a compound composed of two {\it interacting} subsystems. Within the formalism of Brownian motion, it is then possible to introduce a joint-probability for two random forces each acting locally to a subsystem. By contrast, since action is evaluated in configuration space rather than in ordinary space, then in the statistical model based on a random fluctuations of infinitesimal stationary action, one can {\it not} define a joint-probability density for the fluctuations of infinitesimal stationary action of each subsystem.   

We have thus a class of stochastic models which differ fundamentally from the conventional Brownian motion. In the following subsections, we shall select one of them by imposing the constraints that the statistical model has a smooth classical limit in macroscopic regime and respects the principle of Locality demanded by the theory of relativity. 

\subsection{The dynamics of `information' or `uncertainty' with a smooth Macroscopic Classicality}

To develop a statistical description of the stochastic processes, let us denote the joint-probability density that at time $t$ the configuration of the system is $q$ and a random value of $\xi$ is realized as $\Omega(q,\xi;t)$. We would like to find an equation which describes how $\Omega$ changes along an infinitesimally short trajectory $\mathcal{J}(\xi)$. To do this, instead of working directly with $\Omega$, we shall below consider a quantity defined as 
\begin{equation}
I(q;t,\xi)\doteq-\ln\Omega(q,\xi;t). 
\label{information content}
\end{equation}
This quantity is introduced by Shannon as a measure of information content or the degree of uncertainty of an event. Within the context of the stochastic model under study, fixing $\xi$, it is the information content or the degree of uncertainty that the configuration of the system {\it is} $q$ for the following intuitive reasons: i) it is vanishing if the system definitely has a configuration $q$ so that $\Omega(q)=1$; ii) it is increasing monotonically as the probability that the system has a configuration $q$ is decreasing and  iii) it is additive for independent events. 

Let us first note that the information content or the degree of uncertainty defined above is {\it objective} referring directly to the configuration, thus the {\it factual} state, of the system. It is {\it not} the information that one obtains by performing some measurements over the system of interest. Hence, we shall in the present paper work with information within a realist rather than instrumentalist-operational theoretical model. The latter approach is however followed by most works in the reconstruction of quantum mechanics based on information theory, which is apparently motivated by the central role of measurement in the formalism of standard quantum mechanics. Moreover, let us note that the information quantified by $I(q)$ refers to a single event that the system has a particular configuration $q$, rather than the whole possible events of the system distributed according to $\Omega(q)$. The information with regard to the whole possible events is usually quantified by the average of $I(q)$ given by the Gibbs-Shannon entropy which is central in information theory \cite{if}.

The interpretation of $I(q)$ as the amount of information or degree of uncertainty that the system has a configuration $q$ may also be argued within the concept of microcanonical ensemble as follows. First, given the parameters of the system, let $N$ denotes the total number of the microstates accessible by the system. Let us assume that the system may be in one of the microstates equally probably. Let us then assume that $q$ is a `macroscopic coarse-grained variable' of the microstates. Let $N_q$ denotes the number of microstates compatible with $q$. The probability that the system has a configuration $q$ is thus given by $\Omega=N_q/N$. One therefore has $\ln N=I(q)+\ln N_q$. Interpreting $\ln N$ as the amount of information or uncertainty that the system lies in one of the $N$ possible microstates, and $\ln N_q$ as the amount of information or uncertainty that the system lies in one of the $N_q$ microstates compatible with $q$, then it is natural to interpret $I(q)$ as the amount of information or uncertainty that the system has a configuration $q$. 

To avoid confusion, below we shall only use the term `uncertainty' to refer to information content or degree of uncertainty of an event. Let us proceed to again consider two infinitesimally close spacetime points $(q;t)$ and $(q+dq;t+dt)$ such that $\xi$ is constant, connected by an infinitesimally short path $\mathcal{J}(\xi)$. Let us then assume that as the configuration evolves along $\mathcal{J}(\xi)$, the uncertainty that the system has a configuration $q$ also changes according to the following balance equation: 
\begin{equation}
dI(q;t,\xi)=-d\ln\Omega(q,\xi;t)=\Sigma(q;t,\xi)dt,
\label{infinitesimal change of information content}
\end{equation}        
where $\Sigma$ is a function of $q$, $\xi$ and $t$. Our main goal in the present section is then to find a unique functional form of $\Sigma$ and express it in terms of the physical properties of the system, by imposing a set of conceptually simple and physically transparent axioms.  

First, it is instructive to impose the condition of Macroscopic Classicality which demands that in a physical regime corresponding to macroscopic world, one should regain the classical mechanics. Since the deviation from classical mechanics, as assumed in the previous subsection, is due to the fluctuations of infinitesimal stationary action induced by the fluctuations of $\xi$, then in the classical limit of macroscopic regime, such fluctuations must be ignorable. In the macroscopic regime, one must therefore regain the dynamics of ensemble of classical trajectories driven by the deterministic flow of classical velocity field. The infinitesimal change of the uncertainty must in this case solely be given by the flux due to the deterministic classical velocity field. Notice then that the uncertainty should increase if the velocity divergence along the infinitesimally short trajectory $\mathcal{J}(\xi)$ is positive and vice versa. On the other hand, since the system under consideration is closed, then probability has to be conserved. These two conditions combined suggest that in the macroscopic regime whose mathematical formulation will be clarified later, $\Sigma$ on the right hand side of Eq. (\ref{infinitesimal change of information content}) must  reduce to 
\begin{equation}
\Sigma\rightarrow\partial_q\cdot v_c=\theta_c, 
\label{Macroscopic Classicality}
\end{equation} 
where $v_c$ is the classical velocity field which is related to the classical Hamiltonian and the Hamilton's principal function via the kinematic part of the Hamilton equation and the Hamilton-Jacobi equation of (\ref{Hamilton-Jacobi condition}) as 
\begin{equation}
v_c=\frac{\partial H}{\partial p}\Big|_{p=\partial_q A}. 
\label{classical velocity field}
\end{equation}
Indeed, inserting Eq. (\ref{Macroscopic Classicality}) into Eq. (\ref{infinitesimal change of information content}), dividing both sides by $dt$ and taking the limit $dt\rightarrow 0$, one obtains the continuity equation 
\begin{equation}
\partial_t\Omega+\partial_q\cdot\big(\Omega v_c\big)=0, 
\end{equation}
which guarantees the conservation of probability. 

Hence the demand of Macroscopic Classicality suggests that the right hand side of Eq. (\ref{infinitesimal change of information content}) should be given by the following terms: 
\begin{eqnarray}
dI(q;t,\xi)=\theta(q;t,\xi)dt+\sigma(q;t,\xi),\hspace{0mm}\nonumber\\
\mbox{with}\hspace{2mm}\theta\doteq\partial_q\cdot v,\hspace{20mm}
\label{Macroscopic Classicality 0}
\end{eqnarray}
where $v$ is a velocity field which in the classical limit of macroscopic regime must approach $v_c$ 
\begin{equation}
v\rightarrow v_c,
\label{Macroscopic Classicality 1}
\end{equation} 
and $\sigma$ is a function of $q$, $\xi$ and $t$ which must be vanishing in the classical limit
\begin{equation}
\sigma(q;t,\xi)\rightarrow 0. 
\label{Macroscopic Classicality 2} 
\end{equation}
$\sigma(q;t,\xi)$ may thus be regarded as the rate of production of uncertainty along the random path $\mathcal{J}(\xi)$ due to the fluctuations of $\xi$ in microscopic regime. 

From the discussion above, especially the demand that $\sigma$ must be vanishing in the classical limit, it is then natural to assume that $\sigma$ is a function of a quantity that measures the deviation from classical mechanics in microscopic regime due to the fluctuations of $\xi$. To identify such a quantity, let us first assume that $\xi$ is the simplest random variable with two possible values, a binary random variable. Without losing generality let us assume that the two possible values of $\xi$ differ from each other only by their signs, namely one is the opposite of the other, $\xi=\pm|\xi|$. Suppose that both realizations of $\xi$ lead to the same path so that $dA(\xi)=dA(-\xi)$. Since the stationary action principle is valid for both values of $\pm\xi$, then such a model must recover the classical mechanics. Hence, the non-classical behavior must be measured by the difference of $dA(\xi)$ at $\pm|\xi|$, $dA(\xi)-dA(-\xi)$. 

Now let us proceed to assume that $\xi$ may take continuous values. Let us assume that even in this case the difference of the values of $dA$ at $\pm\xi$, 
\begin{equation}
Z(q;t,\xi)\doteq dA(q;t,\xi)-dA(q;t,-\xi)=-Z(q;t,-\xi),  
\end{equation}
measures the non-classical behavior of the stochastic processes, namely the larger the difference, the stronger is the deviation from classical mechanics. $\sigma(q;t,\xi)$ should therefore be a function of $Z(q;t,\xi)$:
\begin{equation}
\sigma=\sigma\big(Z(q;t,\xi)\big). 
\label{Macroscopic Classicality 3}
\end{equation}

For later purpose, let us introduce a new stochastic quantity $S(q;t,\xi)$ so that the differential along the segment of path $\mathcal{J}(\xi)$ is given by 
\begin{equation}
dS(q;t,\xi)=\frac{dA(q;t,\xi)+dA(q;t,-\xi)}{2}=dS(q;t,-\xi).  
\label{infinitesimal symmetry}
\end{equation}
Subtracting $dA(q;t,\xi)$ from both sides, one gets
\begin{eqnarray}
dS(q;t,\xi)-dA(q;t,\xi)=\frac{dA(q;t,-\xi)-dA(q;t,\xi)}{2}\nonumber\\
=-Z(q;t,\xi)/2.\hspace{20mm}
\label{average of classical deviation}
\end{eqnarray} 
$\sigma$ of Eq. (\ref{Macroscopic Classicality 3}) may thus be written as a function of $dS(\xi)-dA(\xi)$
\begin{equation}
\sigma=\sigma\big(dS(\xi)-dA(\xi)\big). 
\label{Macroscopic Classicality 4}
\end{equation}
Note that the assumed universality of the law of physics demands that the functional form of $\sigma$ must be independent from the details of the system of interest: the number of particles, masses, etc. It must only depend on $dS-dA$. 

Let us then express the condition of macroscopic classicality of Eqs. (\ref{Macroscopic Classicality 1}) and (\ref{Macroscopic Classicality 2}) in term of $S$ defined above. Let us first assume that the sign of $\xi$ is fluctuating randomly in a time scale $dt$. Let us then denote the time scale for the fluctuations of $|\xi|$ as $\tau_{\xi}$, and assume that it is much larger than $dt$:   
\begin{equation}
\tau_{\xi}\gg dt. 
\end{equation}
Within a time interval of length $\tau_{\xi}$, the magnitude of $\xi$ is thus effectively  constant while its sign fluctuates randomly. In order for the stochastic system to have a smooth classical limit for all time, then it is necessary that the classical mechanics is recovered in a time interval of length $\tau_{\xi}$ during which the magnitude of $\xi$ is effectively constant while its sign fluctuates randomly. As discussed above, for this binary random variable, the classicality is regained when $dA(\xi)=dA(-\xi)$. In this case, one also has $dS(\xi)=dA(\xi)$ by the virtue of Eq. (\ref{average of classical deviation}), so that due to Eq. (\ref{infinitesimal stationary action}), $S$ satisfies the Hamilton-Jacobi equation of (\ref{Hamilton-Jacobi condition}). Taking into account this fact, first, the condition of Macroscopic Classicality of Eq. (\ref{Macroscopic Classicality 2}) should be rewritten as 
\begin{equation}
\lim_{dS\rightarrow dA}\sigma(dS-dA)=0. 
\label{Macroscopic Classicality 5}
\end{equation}
Moreover, to attain the condition of Macroscopic Classicality of Eq. (\ref{Macroscopic Classicality 1}) it is sufficient to assume that $v$ in Eq. (\ref{Macroscopic Classicality 0}) is related to $S$ as follows 
\begin{equation}
v=\frac{\partial H}{\partial p}\Big|_{p=\partial_qS}. 
\label{Macroscopic Classicality 6}
\end{equation}
One can see that in the limit $dS\rightarrow dA$ one has $\partial_qS\rightarrow\partial_qA$ so that $v\rightarrow v_c$ as expected. Let us emphasize that the above condition is sufficient to recover the classical mechanics only within the time interval of length $\tau_{\xi}$ in which $|\xi|$ is constant. While it is also a necessary condition to recover the classical dynamics for the whole time, it is not sufficient. One needs to have more conditions to recover classical mechanics for the whole time. 

\subsection{An infinitesimal change of uncertainty respecting the principle of Locality}

Let us then proceed to show that imposing the principle of Locality will pick up uniquely, up to a free parameter, the functional form of $\sigma(dS-dA)$. To do this, let us consider a compound of two subsystems, say two particles whose configuration are denoted respectively by $q_1$ and $q_2$. Let us assume that they are spacelike separated from each other so that due to the principle of Locality, there is no mechanical interaction between the two particles. The total Lagrangian is thus decomposable as $L(q_1,q_2,\dot{q}_1,\dot{q}_2)=L_1(q_1,\dot{q}_1)+L_2(q_2,\dot{q}_2)$ and accordingly, $dA(q_1,q_2)$ and $dS(q_1,q_2)$ are also decomposable: $dA(q_1,q_2)=dA_1(q_1)+dA_2(q_2)$ and $dS(q_1,q_2)=dS_1(q_1)+dS_2(q_2)$. $\sigma$ can thus be written as 
\begin{equation}
\sigma(dS-dA)=\sigma\big((dS_1-dA_1)+(dS_2-dA_2)\big). 
\label{information production noninteracting}
\end{equation} 
Further, in this case, the classical Hamiltonian is also decomposable $H(q_1,q_2,p_1,p_2)=H_1(q_1,p_1)+H_2(q_2,p_2)$, where $p_i$, $i=1,2$, is the momentum of $i-$particle. Putting this into Eq. (\ref{Macroscopic Classicality 6}) and recalling that $dS$ is decomposable, then $\theta$ defined in Eq. (\ref{Macroscopic Classicality 0}) is also decomposable 
\begin{equation}
\theta(q_1,q_2)=\theta_1(q_1)+\theta_2(q_2).
\label{decomposability 2}
\end{equation} 
The change of the uncertainty that the compound system has a configuration $q=(q_1,q_2)$ moving along the path $\mathcal{J}(\xi)$ thus reads, by the virtue of Eq. (\ref{information production noninteracting}) and (\ref{decomposability 2}),
\begin{equation}
dI(q_1,q_2)=\big(\theta_1+\theta_2\big)dt+\sigma\big((dS_1-dA_1)+(dS_2-dA_2)\big).
\label{LIC non-interacting compound system}
\end{equation}

On the other hand, since the two subsystems are spacelike separated from each other, the principle of Locality demands that the change of the uncertainty that the first (second) subsystem has a configuration $q_1$ ($q_2$), when the compound system moves along an infinitesimally short trajectory $\mathcal{J}(\xi)$, must be independent from what happens with the second (first) subsystem. Otherwise, the uncertainty that one subsystem has a certain configuration can be influenced by the state of the other distantly separated subsystem by varying the control parameters of the latter despite of no interaction. Hence, the change of the uncertainty that each subsystem has a certain configuration must only depend on the corresponding single particle Lagrangian. One therefore has the following pair of decoupled relations:
\begin{eqnarray}
dI_1(q_1)=\theta_1dt+\sigma(dS_1-dA_1),\nonumber\\
dI_2(q_2)=\theta_2dt+\sigma(dS_2-dA_2),
\label{LIC single particle non-interacting}
\end{eqnarray}
where $dI_i=-d(\ln\Omega_i)$, and $\Omega_i(q_i)$, $i=1,2$, is the probability density for the configuration of the $i-$particle. $dI_i$ is thus the change of the uncertainty that the $i-$subsystem has a configuration $q_i$. Let us note again that the assumed universality of the law of physics demands that the functional form of $\sigma$ for the whole compound system on the right hand side of Eq. (\ref{LIC non-interacting compound system}) must be the same as those for each subsystem on the right hand side of Eq. (\ref{LIC single particle non-interacting}). 

Let us first assume that the probability distribution of the configuration of the compound system is separable: $\Omega(q_1,q_2)=\Omega_1(q_1)\Omega_2(q_2)$. In this case, the total change of the uncertainty that the compound system has a configuration $q=(q_1,q_2)$ as the configuration evolves along $\mathcal{J}(\xi)$ is then decomposable as $dI(q_1,q_2)=dI_1(q_1)+dI_2(q_2)$. Now let us consider a general case when the distribution of the configuration of the two spacelike separated subsystems are correlated. One thus has $\Omega(q_1,q_2)=\Omega_{12}(q_1|q_2)\Omega_2(q_2)$, where $\Omega_{12}(q_1|q_2)$ is the conditional probability that the configuration of the first subsystem is $q_1$ when the configuration of the second subsystem is $q_2$. As the configuration of the compound system evolves along an infinitesimally short path $\mathcal{J}(\xi)$, the total change of the uncertainty that the compound system has a configuration $q=(q_1,q_2)$ is then 
\begin{equation}
dI(q_1,q_2)=dI_{12}(q_1|q_2)+dI_2(q_2), 
\label{change of information content general}
\end{equation}
where $dI_{12}=-d\ln\Omega_{12}$ is the infinitesimal change of the uncertainty that the configuration of the first subsystem is $q_1$ when the configuration of the second subsystem is $q_2$. The principle of Locality however demands that, since the two subsystems are spacelike separated from each other, the infinitesimal change of the uncertainty that the first subsystem has a configuration $q_1$ must be independent of the configuration of the second subsystem $q_2$. One must thus have $dI_{12}(q_1|q_2)=dI_1(q_1)$. Inserting into Eq. (\ref{change of information content general}), one therefore concludes that in general the total infinitesimal change of the uncertainty that the two non-interacting subsystems have a configuration $q=(q_1,q_2)$ is decomposable as
\begin{equation}
dI(q_1,q_2)=dI_1(q_1)+dI_2(q_2).
\label{decomposability 1} 
\end{equation}

Finally inserting Eqs. (\ref{LIC non-interacting compound system}) and (\ref{LIC single particle non-interacting}) into Eq. (\ref{decomposability 1}), in general $\sigma(dS-dA)$ must also satisfy the following decomposability condition: 
\begin{equation}
\sigma\big((dS_1-dA_1)+(dS_2-dA_2)\big)=\sigma(dS_1-dA_1)+\sigma(dS_2-dA_2). 
\label{principle of Local-Causality}
\end{equation} 
The above functional equation together with the necessary condition of Macroscopic Classicality of Eq. (\ref{Macroscopic Classicality 5}) can then be solved to give the following linear solution:
\begin{equation}
\sigma(dS-dA)=\alpha(\xi;t)(dS-dA),
\label{Local-Causality 1} 
\end{equation}
where $\alpha$ is a real-valued function independent from $dS-dA$, yet might depend on $t$ and $\xi$, hence is randomly fluctuating with $\xi$ in a microscopic time scale. Let us emphasize that Eq. (\ref{Local-Causality 1}) now applies for general cases, not only for a compound of non-interacting subsystems. 

For the reason that will be clear later, let us introduce a new non-vanishing random variable $\lambda(\xi;t)=2/\alpha$. The change of the uncertainty that the system has a certain configuration along an infinitesimally short path $\mathcal{J}(\xi)$ of Eq. (\ref{infinitesimal change of information content}) is thus given by 
\begin{equation}
dI=-d\ln\Omega=\theta dt+\frac{2}{\lambda}(dS-dA).  
\label{fundamental equation}
\end{equation}  
Further, since $\xi$ is fixed during the time interval $dt$, one can expand all the differentials as $dF=\partial_tF dt+\partial_qF\cdot dq$ to have the following pair of coupled partial differential equations:
\begin{eqnarray}
-\partial_q\ln\Omega=\frac{2}{\lambda}\Big(\partial_qS-p(\dot{q})\Big),\hspace{8mm}\nonumber\\
-\partial_t\ln\Omega=\frac{2}{\lambda}\Big(H(q,p)+\partial_tS\Big)+\theta(S), 
\label{fundamental equation 1}
\end{eqnarray} 
where we have made use of Eq. (\ref{infinitesimal stationary action}). The spatial and temporal changes of the uncertainty that the system has a certain configuration are thus related to the momentum and energy of the system, respectively. Let us emphasize that the above pair of equations are valid only for a time interval in which $\xi$ is constant. 

Let us write the pair of coupled equations of Eq. (\ref{fundamental equation 1}) as follows:
\begin{eqnarray}
p(\dot{q})=\partial_qS+\frac{\lambda}{2}\frac{\partial_q\Omega}{\Omega},\hspace{8mm}\nonumber\\
-H(q,p)=\partial_tS+\frac{\lambda}{2}\frac{\partial_t\Omega}{\Omega}+\frac{\lambda}{2}\theta(S), 
\label{fundamental equation rederived}
\end{eqnarray} 
where the momentum and energy are put on the left hand side. The above pair of relations must {\it not} be interpreted that the momentum or velocity and energy of the system are determined {\it causally} by the change of the uncertainty, which is physically absurd. Rather both the momentum and energy provide the source of change of the uncertainty that the system has a certain configuration along an infinitesimally short trajectory $\mathcal{J}(\xi)$ as shown explicitly by Eq. (\ref{fundamental equation}). Further, it is evident that in the formal limit $\lambda\rightarrow 0$ whose physical meaning will be clarified in the next subsection, Eq. (\ref{fundamental equation rederived}) reduces back to the Hamilton-Jacobi equation of (\ref{Hamilton-Jacobi condition}). In this sense, Eq. (\ref{fundamental equation rederived}) can be regarded as a generalization of the Hamilton-Jacobi equation. Unlike the Hamilton-Jacobi equation in which we have a single unknown function $A$, however, to calculate the velocity or momentum and energy, one now needs a pair of unknown functions $S$ and $\Omega$.  

\subsection{A stochastic processes with a transition probability given by an exponential distribution of deviation from infinitesimal stationary action}

We have started from a stochastic processes in which the system with a configuration $q$ at time $t$ can take one of many possible random paths $\mathcal{J}(\xi)$ selected by the principle of stationary action with different random values of $\xi$, to end up with a configuration $q+dq$ at time $t+dt$. We then derived a law of infinitesimal change of the uncertainty of an event along an infinitesimally short path by imposing the condition of Macroscopic Classicality and the principle of Locality. It is then tempting to investigate if the law of change of uncertainty given by Eq. (\ref{fundamental equation}) completely determines the stochastic processes. To see this, fixing $\xi$, let $\Omega\big(\{(q+dq;t+dt),(q;t)\}\big|\mathcal{J}(\xi)\big)$ denotes the conditional joint-probability density that the configuration of the system is $q$ at time $t$, tracing the trajectory $\mathcal{J}(\xi)$ and end up with a configuration $q+dq$ at time $d+dt$. Using this quantity, the change of probability density $d\Omega$ due to the transport along the path $\mathcal{J}(\xi)$ is given by 
\begin{equation}
d\Omega(q,\xi;t)=\Omega\big(\{(q+dq;t+dt),(q;t)\}\big|\mathcal{J}(\xi)\big)-\Omega(q,\xi;t). 
\end{equation}
Inserting into Eq. (\ref{fundamental equation}) one therefore has 
\begin{eqnarray}
\Omega\big(\{(q+dq;t+dt),(q;t)\}\big|\mathcal{J}(\xi)\big)\hspace{10mm}\nonumber\\
=\big[1-\theta(S)dt-\frac{2}{\lambda}(dS-dA)\big]\Omega(q,\xi;t).
\label{pre transition probability}
\end{eqnarray}

Let us then consider the case when $|(dS-dA)/\lambda|\ll 1$. Equation (\ref{pre transition probability}) can then be written approximately as   
\begin{eqnarray}
\Omega\Big(\{(q+dq;t+dt),(q;t)\}\big|\mathcal{J}(\xi)\Big)\hspace{10mm}\nonumber\\
\approx e^{-\frac{2}{\lambda}(dS(\xi)-dA(\xi))-\theta(S) dt}\times\Omega(q,\xi;t).  
\label{probability density} 
\end{eqnarray}
The above relation can obviously be read within the conventional probability theory as follows: the joint-probability density that the system initially at $(q;t)$ traces the segment of trajectory $\mathcal{J}(\xi)$ and end up at $(q+dq;t+dt)$, $\Omega\big(\{(q+dq;t+dt),(q;t)\}\big|\mathcal{J}(\xi)\big)$, is equal to the probability that the configuration of the system is $q$ at time $t$, $\Omega(q,\xi;t)$, multiplied by a `transition probability' between the two infinitesimally close spacetime points via the segment of trajectory $\mathcal{J}(\xi)$ given by 
\begin{equation}
P((q+dq;t+dt)|\{\mathcal{J}(\xi),(q;t)\})\propto e^{-\frac{2}{\lambda}(dS(\xi)-dA(\xi))}/\mathcal{Z},
\label{exponential distribution of DISA}  
\end{equation}
where $\mathcal{Z}=\exp(\theta(S) dt)$.  

Some notes are instructive. First, to guarantee the normalizability of the above transition probability, then the exponent $(dS(\xi)-dA(\xi))/\lambda(\xi)$ must be non-negative for any spacetime point $(q,t)$. This demands that $dS(\xi)-dA(\xi)$ must always have the same sign as $\lambda(\xi)$. On the other hand, from Eq. (\ref{average of classical deviation}), one can see that $dS(\xi)-dA(\xi)$ changes its sign as $\xi$ flips its sign. Hence, to guarantee the non-negativity of $(dS(\xi)-dA(\xi))/\lambda(\xi)$, $\lambda$ must also change its sign as $\xi$ flips its sign. This allows us to assume that the sign of $\lambda$ is always the same as that of $\xi$. The time scale for the fluctuations of the sign of $\lambda$ must therefore be the same as the time scale for the fluctuations of $\xi$ given by $dt$. Moreover, the second term on the right hand side of Eq. (\ref{fundamental equation}), $\sigma$, namely the rate of production of uncertainty due to the fluctuations of infinitesimal stationary action is always non-negative.  

It is also evident that for the distribution of Eq. (\ref{exponential distribution of DISA}) to make sense mathematically, the time scale for the fluctuations of $|\lambda|$, denoted by $\tau_{\lambda}$, must be much larger than that of $|\xi|$. One thus has  
\begin{equation}
\tau_{\lambda}\gg\tau_{\xi}\gg dt.  
\label{time scales}
\end{equation}
In other words, $|\xi|$ fluctuates much faster than $|\lambda|$, yet both $\xi$ and $\lambda$ always have the same sign fluctuating randomly in the time scale $dt$. Hence, within a time interval of length $\tau_{\lambda}$ during which $|\lambda|$ is effectively constant, one may assume that $dS(\xi)-dA(\xi)$ is randomly fluctuating due to the fluctuations of $|\xi|$ distributed according to the exponential law of Eq. (\ref{exponential distribution of DISA}) characterized by $|\lambda|$. 

Next, there is no a priori reason on how the sign of the values of $dS-dA$ should be distributed. Following the principle of indifference (principle of insufficient reason) \cite{Jaynes book}, it is then advisable to assume that the sign of $dS-dA$ is distributed equally probably. Further, since the sign of $dS(\xi)-dA(\xi)$ changes as $\xi$ flips its sign, then the sign of $\xi$ must also be fluctuating randomly with equal probability so that the probability density of the value of $\xi$ at any given time, denoted below by $P_{H}(\xi)$, must satisfy the following unbiased condition: 
\begin{equation}
P_{H}(\xi)=P_{H}(-\xi). 
\label{God's unbiased}   
\end{equation}
Since the sign of $\lambda$ is always the same as that of $\xi$ then the probability distribution function of $\lambda$ must also satisfy the same unbiased condition. Further, since $P_{H}(\xi)=\int dq\Omega(q,\xi;t)$, then Eq. (\ref{God's unbiased}) demands the following symmetry relation:  
\begin{eqnarray} 
\Omega(q,\xi;t)=\Omega(q,-\xi;t). 
\label{God's fairness}
\end{eqnarray}   
One also has, from Eq. (\ref{infinitesimal symmetry}), the following symmetry relations for the spatiotemporal gradient of $S(q,\xi;t)$:
\begin{eqnarray}
\partial_qS(q;t,\xi)=\partial_qS(q;t,-\xi),\nonumber\\
\partial_tS(q;t,\xi)=\partial_tS(q;t,-\xi).
\label{quantum phase symmetry}
\end{eqnarray} 

Recall that the pair of relations in Eqs. (\ref{fundamental equation 1}) or (\ref{fundamental equation rederived}) are valid when $\xi$ is fixed. However, since as discussed above, the second term on the right hand side of Eq. (\ref{fundamental equation}) is insensitive to the sign of $\lambda$ which is always the same as the sign of $\xi$, then the pair of equations in (\ref{fundamental equation rederived}) are valid in a microscopic time interval of length $\tau_{\xi}$ during which the magnitude of $\xi$ is constant while its sign may change randomly. To have an evolution for a finite time interval $t>\tau_{\xi}$, one can proceed along the following approximation. First one divides the time into a series of microscopic time intervals of length $\tau_{\xi}$: $t\in[(k-1)\tau_{\xi},k\tau_{\xi})$, $k=1,2,\dots$, and attributes to each interval a random value of $\xi(t)=\xi_k$ according to the probability distribution $P_{H_k}(\xi_k)=P_{H_k}(-\xi_k)$. Hence, during the interval $[(k-1)\tau_{\xi},k\tau_{\xi})$, the magnitude of $\xi(t)=\xi_k$ is kept constant while its sign changes randomly in an infinitesimal time scale $dt$. One then applies the pair of relations in Eqs. (\ref{fundamental equation 1}) or (\ref{fundamental equation rederived}) during each interval of time with fixed $|\xi(t)|=|\xi_k|$, consecutively.

Since $dA$ is just the infinitesimal stationary action along the short path $\mathcal{J}(\xi)$, $|dS-dA|$ may be regarded as the deviation from infinitesimal stationary action, the distribution of which is given by Eq. (\ref{exponential distribution of DISA}). Such an exponential distribution was firstly suggested heuristically in Ref. \cite{AgungSMQ4} to model a microscopic stochastic deviation from classical mechanics. An application of the statistical model to model quantum measurement is given recently in Ref. \cite{AgungSMQ7}. For a fixed value of $|\lambda|$ which is valid during a time interval of length $\tau_{\lambda}$, one can see from Eq. (\ref{exponential distribution of DISA}) that the average deviation from infinitesimal stationary action is given by  
\begin{equation}
\overline{|dS-dA|}=|\lambda|/2. 
\label{average deviation from infinitesimal action} 
\end{equation}
It is then evident that in the regime where the average deviation is much smaller than the infinitesimal stationary action itself, namely $|dA/\lambda|\gg 1$, or formally in the limit $|\lambda|\rightarrow 0$, Eq. (\ref{exponential distribution of DISA}) reduces to 
\begin{equation}
\rightarrow\delta(dS-dA),
\end{equation}
so that $dS(\xi)\rightarrow dA(\xi)$. Such a regime thus must be identified as the macroscopic regime. This fact suggests that $|\lambda|$ must have a very small microscopic value. In this regard, the pair of equations in (\ref{fundamental equation rederived}) may be regarded as a generalization of the Hamilton-Jacobi equation of (\ref{Hamilton-Jacobi condition}) due to the exponential distribution of deviation from infinitesimal stationary action of Eq. (\ref{exponential distribution of DISA}). Let us also note that since $|\lambda|$ in general may depend on time, then the transition probability is in general {\it not} stationary except when $\lambda=\pm\mathcal{Q}$ all the time, where $\mathcal{Q}$ is a constant. 

One can also see that the decomposability of the infinitesimal change of the uncertainty for a pair of spacelike separated subsystems given by Eq. (\ref{decomposability 1}), which is demanded by the principle of Local-Causality, implies directly the separability of the transition probability of Eq. (\ref{exponential distribution of DISA}) for the non-interacting subsystems. Namely, for non-interacting two subsystems such that $dA$ and $dS$ are decomposable as $dA(q_1,q_2)=dA_1(q_1)+dA_2(q_2)$ and $dS(q_1,q_2)=dS_1(q_1)+dS_2(q_2)$, respectively, so that $\theta$ is also decomposable $\theta(q_1,q_2)=\theta_1(q_1)+\theta_2(q_2)$, the transition probability of Eq. (\ref{exponential distribution of DISA}) is separable as 
\begin{equation}
P_S(dS_1+dS|dA_1+dA_2)=P_S(dS_1|dA_1)P_S(dS_2|dA_2). 
\label{separability condition}
\end{equation}
Hence, the transition probability which determines the stochastic behavior of one subsystem is independent from that of the other subsystem as intuitively expected for spacelike separated non-interacting subsystems. Otherwise, the dynamics and statistics of one subsystem is influenced by the other subsystem, which contradicts the principle of Locality. See also Ref. \cite{AgungSMQ6} for a different approach to single out Eq. (\ref{exponential distribution of DISA}) by imposing the principle of Locality without directly employing the concept of information or uncertainty.

\section{Quantization}

We have shown in Refs. \cite{AgungSMQ4,AgungSMQ6,AgungSMQ7}, that applying the above statistical model to a wide class of classical systems, then the dynamics of the ensemble of trajectories is governed in the lowest order approximation by the Schr\"odinger equation with Born's statistical interpretation of wave function reproducing the formal results of canonical quantization, if $\lambda=\pm\hbar$ for all time so that the average deviation from infinitesimal stationary action distributed according to the exponential law of Eq. (\ref{exponential distribution of DISA}) is given by $\hbar/2$. This is done by identifying the wave function as 
\begin{equation}
\Psi\doteq \sqrt{\Omega}\exp\Big(i\frac{S}{|\lambda|}\Big). 
\label{general wave function}
\end{equation}
The statistical model also leads to an objective uncertainty relation which furthermore implies the quantum mechanical canonical uncertainty relation. For a related work, see also Ref. \cite{AgungSMQ3}.  

The abstract rules of canonical quantization thus `effectively' arise from a statistical modification of classical mechanics in microscopic regime based on a specific law of change of the uncertainty (information) that the system has a certain configuration of Eq. (\ref{fundamental equation}) chosen uniquely by imposing the condition of Macroscopic Classicality and the principle of Local-Causality. Unlike the canonical quantization which is formal-mathematical with obscure physical meaning, the statistical model of quantization is thus `physical'. Moreover, Planck constant acquires a physical interpretation as the average deviation from classical mechanics in a microscopic time scale. Further, unlike canonical quantization which in general leads to an infinite number of possible quantum Hamiltonians for a single classical Hamiltonian which is due to the replacement of c-numbers (classical numbers) by q-numbers (quantum numbers/Hermitian operators), since the statistical model is based on a manipulation of c-numbers, it always gives a unique quantum Hamiltonian with a specific ordering of operators \cite{AgungSMQ4,AgungSMQ6}. 

Recall that in standard quantum mechanics, the state of the system is assumed to be determined completely by specifying the wave function. The wave function is thus regarded as fundamental. Statements about position and momentum are then relegated {\it operationally} to certain acts of measurement over the state of the system represented by the wave function. The canonical uncertainty relation between the statistical results of position and momentum measurement is usually mentioned to support the argumentation that it is impossible to attribute a pair of definite values of position and momentum to a system, nor is such an attribution useful. By contrast, in the statistical model of quantization developed in the present paper, one assumes the objective ontology of particles with a definite configuration for all the time as in classical mechanics. Hence, configuration of the system is regarded as the ``beable'' of the theory in Bell's sense \cite{Bell beable}. The wave function, on the other hand, is argued to be emergent artificial convenient mathematical tool to describe the dynamics and statistics of ensemble of trajectories. The objective ontology of the trajectories guarantees a conceptually smooth classical correspondence. 

One can also show that the actual trajectory of the particle is in general fluctuating randomly around the so-called Bohmian trajectory in pilot-wave theory \cite{Bohmian trajectory}. Hence, we have a physical picture that the actual trajectory is fluctuating randomly around the Bohmian trajectory while the latter moves {\it as if} it is guided by the wave function evolving deterministically according to the Schr\"odinger equation. Yet, unlike the pilot-wave theory, the wave function in the statistical model is {\it not} physically real but an artificial mathematical construct, and the Schr\"odinger equation and the guidance relation are derived from first principle rather than ad-hoc-ly postulated as in pilot-wave theory. Recall that the fundamental assumption in pilot-wave theory that the wave function is a physical field, living in configuration space rather than in ordinary space, is known to lead to a conceptual difficulty, and furthermore implies rigid nonlocality in direct conflict with the special theory of relativity. By contrast, the present statistical model is developed based on the principle of Locality. In this sense, the upper equation in (\ref{fundamental equation rederived}) can not be regarded as a causal-dynamical guidance relation as in pilot-wave theory, but a kinematical relation. Moreover, unlike the pilot-wave theory which is deterministic and relegates the microscopic randomness to our ignorance of the initial condition, the statistical model is strictly stochastic. 

Notice that as argued in the previous section, the unique form of the law of infinitesimal change of the degree of uncertainty that the system has a certain configuration along a short path given by Eq. (\ref{fundamental equation}) is singled out by imposing the condition of Macroscopic Classicality and the principle of Locality. In particular, the principle of Locality is decisive in selecting the linear form of the second term on the right hand side of Eq. (\ref{fundamental equation}) which describes the production of information or uncertainty due to the fluctuation of infinitesimal stationary action. Since the stochastic processes based on such a change of the information leads to the derivation of the {\it linear} Schr\"odinger equation, one may thus argue that the principle of Locality expressed in Eq. (\ref{principle of Local-Causality}) is a necessary condition for the linearity of the Schr\"odinger equation \cite{AgungSMQ6}. To support this argumentation, let us mention that a nonlinear extension of quantum dynamics \cite{Weinberg nonlinearity} may lead to signaling \cite{Gisin superluminal signaling,Polchinski superluminal signaling} thus violating the principle of Locality. Similarly, as argued in Refs. \cite{AgungSMQ3,AgungSMQ6}, since the quantum mechanical uncertainty relation can be derived from the upper equation in (\ref{fundamental equation rederived}), then the principle of Locality is also necessary for the former. 

Finally, we have also shown in Ref. \cite{AgungSMQ6} for a system of spin-less particles that the average of the physically relevant quantities over the distribution of the configuration are equal to the quantum mechanical averages of the corresponding quantum mechanical Hermitian operators over a quantum state. Note however that while the former refers to the average of the actual values of the physical quantities, the latter refers to the statistical average of measurement outcomes in an ensemble of identical experiments. This result applies to any physical quantity of a function of position and momentum with up to second degree of the latter. An application of the statistical model to model quantum mechanical measurement of angular momentum without wave collapse, reproducing the prediction of quantum mechanics is reported in Ref. \cite{AgungSMQ7}. 

\section{Conclusion and Remarks} 

We first developed a stochastic processes for a microscopic stochastic deviation from classical mechanics in which the randomness is modeled by a stochastic fluctuations of the infinitesimal stationary action, thus is physically different from that of the Brownian motion based on random forces. Such a stochastic processes leads to a production of the uncertainty that the system has a certain configuration in a microscopic time scale, which is assumed to be vanishing in the classical limit of macroscopic regime. We then showed that imposing the principle of Locality, which requires the infinitesimal change of the uncertainty that a subsystem has a certain configuration to be independent from the configuration of the other spacelike separated subsystem, will select a unique law of infinitesimal change of uncertainty, up to a free parameter. We further showed that such a law of infinitesimal change of uncertainty determines a stochastic processes with a transition probability between two infinitesimally close spacetime points along a randomly chosen path that is given by an exponential distribution of deviation from infinitesimal stationary action. 

Given a classical Hamiltonian, we have shown in the previous works \cite{AgungSMQ4,AgungSMQ6,AgungSMQ7} that the statistical model leads to the derivation of Schr\"odinger equation with Born's statistical interpretation of wave function. The model also leads to an objective uncertainty relation which implies the standard quantum mechanical uncertainty relation. Unlike the canonical quantization, however, in the statistical model, the system always has a definite configuration all the time as in classical mechanics, fluctuating randomly along a continuous trajectory. We have also shown, for a system of spin-less particles, that the average of the relevant physical quantities over the distribution of the configuration is numerically equal to the quantum mechanical average of the corresponding quantum mechanical Hermitian operators over a quantum state represented by a wave function. Since the principle of Locality is derived from our conception of spacetime structure, then one may conclude that the dynamics and kinematics of quantum mechanics is intimately related to the former.  

Some problems are left for future study. It is first imperative to ask how such a local-causal statistical model would explain the violation of Bell's inequalities predicted by the quantum mechanics and verified in numerous experiments which is widely believed to give a strong evidence that Nature is nonlocal? This is a crucial problem needed an explanation within the statistical model. Recall that Bell's inequalities are derived by assuming 1) the separability of probability of outcomes in a pair of spacelike joint-measurements (Bell's locality assumption) and 2) the so-called measurement independence or experimental free-will, that the distribution of the hidden variables underlying the measurement outcomes is independent from the setting parameters of the apparatus chosen freely by the observer \cite{Bell free will,Shimony free will,Espagnat free will,Hall free will relaxation}. It is tempting to guess that the objective locality of the statistical model implies the Bell's locality assumption so that the model must somehow violate measurement independence. It is therefore instructive to study the above two fundamental hypothesis within the statistical model by first applying the model to develop quantum measurement in realistic physical systems and derive the Born's rule \cite{AgungSMQ7}.  
 
Next, it is also tempting to ask why canonical quantization corresponds to a specific case when $|\lambda|$ in Eq. (\ref{fundamental equation}), the free parameter of the statistical model, is given by $\hbar$ so that the average deviation from infinitesimal stationary action distributed according to the exponential law of Eq. (\ref{exponential distribution of DISA}) is given by $\hbar/2$. Is the relation $|\lambda|=\hbar$ exact? Or whether Nature allows for a small fluctuations of $|\lambda|$ around $\hbar$? Recall also that the Schr\"odinger equation is derived as the zeroth order approximation of the statistical model. It is then imperative to study the higher orders corrections. These last two cases may thus provide precision tests against quantum mechanics.        

\begin{acknowledgments} 

\end{acknowledgments}

\end{document}